\documentstyle[preprint,prb,aps]{revtex}
\begin{document}

\title{Quantum Hall - insulator transitions in lattice models with 
strong disorder}

\author{Kun Yang}
\address{Department of Electrical Engineering, Princeton University,
Princeton, NJ 08544\\
and Condensed Matter Physics 114-36,
California Institute of Technology,
Pasadena, CA 91125}
\author{R. N. Bhatt}
\address{Department of Electrical Engineering, Princeton University,
Princeton, NJ 08544}
\date{\today}

\maketitle

\begin{abstract}
We report results of numerical studies of the integer quantum Hall effect
in a tight binding model 
on a two-dimensional square lattice with non-interacting electrons, in the
presence of a random potential 
as well as a uniform magnetic field applied perpendicular to the lattice.
We consider field magnitudes such 
that the area per flux quantum is commensurate with the lattice
structure. 
Topological properties of the single electron wave functions are used to 
identify current carrying states that are responsible for the quantized 
Hall conductance. We study the interplay between the magnetic field and 
the disorder, and find a universal pattern with which the current carrying
states are destroyed by increasing disorder strength, and the system driven
into an insulating state. We also discuss how to interpolate results of lattice
models to the continuum limit.
The relationship to previous 
theoretical and experimental studies of quantum Hall-insulator transitions
in strongly disordered systems at low magnetic fields
is discussed. 

\end{abstract}


\section{Introduction}
\label{sec:intro}

The integer quantum Hall effect (IQHE) is observed in two-dimensional (2D)
electron systems, in the presence of a strong perpendicular 
magnetic field.\cite{pg} Current understanding of IQHE is mostly based on 
studies of non-interacting electron models, with the simplifying
assumption that electron-electron interaction does not play an important
role. Indeed, the phenomenology of IQHE may be explained very well by 
the non-interacting electron model in a strong magnetic field:
the magnetic field gives rise to Landau quantization and the large
gap necessary for dissipationless transport, 
while disorder localizes electronic states away
from the centers of Landau bands, such that the Hall conductance
remains quantized as the Fermi energy lies in the localized region of the
density of states (DOS); transitions between different quantum Hall phases
occur when the Fermi energy sweeps through critical energies in the
middle of Landau bands, where states are delocalized (i.e., have a diverging localization length).\cite{bodoreview,noteadd}

The scaling theory of localization\cite{gangof4} predicts that
all states in a system of noninteracting electrons are localized in 2D, in the absence of 
a magnetic field. Therefore a 2D electron gas system must undergo
a phase transition (or a series of phase transitions) from a quantum Hall 
state to the insulating phase, as the strength of the magnetic field ($B$)
goes to zero.\cite{note} The nature of such transition(s) is of current 
interest,\cite{jiang,wang,shahar,song}
and is the subject of the present paper.
On the theoretical front, Khmelnitskii\cite{kh} and Laughlin\cite{laughlin}
proposed, based on semiclassical considerations, that
delocalized states at centers of Landau bands ``float" upward as 
$B$ decreases and mixing between different Landau bands becomes 
important, and disappear at infinite
energy in the $B\rightarrow 0$ limit, thus reconciling the very different
behavior at zero and strong magnetic fields.
This simple and elegant picture is the basis of the global phase diagram
of quantum Hall effect proposed by Kivelson, Lee and Zhang,\cite{klz}
and has received support from both 
experimental\cite{shashkin,glozman,krav,furneaux} and 
microscopic theoretical\cite{a,ld,shahbazyan,yang,gramada,haldane}
studies. 
On the other hand, based on numerical studies of a one-band
tight binding model on a square lattice, authors of Refs. \onlinecite{liu,xie}
(see also Refs. \onlinecite{sheng,sheng2})
concluded that instead of floating to infinite energy in the 
$B\rightarrow 0$ limit, the extended states
disappear at finite $B$, and proposed a very different phase diagram.
On the experimental side, the situation is also rather unclear.
Although most experiments are consistent with the global phase diagram,
people have routinely observed direct transition from quantum Hall 
phases with $\nu > 1$ to an insulator as $B$ decreases, where $\nu$ is the
Hall conductance in unit of ${e^2\over h}$. This is in apparent 
contradiction with the global phase diagram, which predicts that $\nu$ can
only change by $\pm 1$ at each integer quantum Hall transition.
Some of these discrepancies, like the $2\rightarrow 0$ 
transitions,\cite{jiang,wang}
may be explained as due to the small 
Zeeman splitting at low $B$, which gives rise to spin-unresolved quantum Hall
transitions.
The recently observed $3\rightarrow 0$ transition,\cite{song} however,
is apparently quite difficult to interpret along the same line.

Most of the previous theoretical studies focus on the continuum model.
It was recently argued,\cite{liu,xie,sheng,sheng2} however, that the
existence of an underlying lattice (and the corresponding conduction band)
is important to understand the 
transition between a quantum Hall state and the insulating state at low
$B$, even though the band width is much bigger than the Fermi energy and 
Landau level spacings. 
A crucial difference between the continuum model and tight-binding 
models is that in the latter case, there exist current carrying states that
carry negative Hall conductance near the band center (and the total Hall
conductance is zero for the entire band), while such states do not exist 
in the former case. However 
it remains unclear how to interpolate between the lattice
and continuum models.

In this paper we present new results of
numerical studies of the tight binding lattice model in the presence of
a uniform magnetic field and random potential. 
Unlike previous 
studies\cite{yang,liu,xie,sheng,sheng2,ando,tan}
which have been focusing on special cases where there is particle-hole
symmetry and critical energies carrying negative Hall conductance
only exist in the central band(s), we have studied the
most general case. We find, however, that the pattern by which the extended states 
disappear as disorder strength increases is rather similar to that of the
special case studied earlier. 
We also explicitly discuss how to interpolate the numerical results
on a lattice to the continuum limit properly.

The rest of the paper is organized in the following way.
In section \ref{model} we introduce the model we study and the numerical 
approach we use to identify extended current carrying states, namely the
calculation of a topological quantum number called Chern number. In 
section \ref{numer} we present our numerical results for various cases that
we have studied. In section \ref{sum} we summarize our results, and
discuss their relations with previous theoretical and experimental studies. 

\section{Model and chern numbers}
\label{model}

We study numerically non-interacting electrons moving on a 
lattice with a uniform magnetic field and random onsite potential, 
described by the following tight binding Hamiltonian:
\begin{equation}
H=-\sum_{ij}t_{ij}(e^{ia_{ij}}c_i^{\dagger}c_j+
e^{-ia_{ij}}c_j^{\dagger}c_i)+\sum_i\epsilon_ic_i^{\dagger}c_i,
\label{ham}
\end{equation}
where $c_i$ is the fermion operator on lattice site $i$.
The first term
represents hopping or kinetic energy of the electrons,
and the hopping phases $a_{ij}$ are determined by the uniform magnetic 
field (up to arbitrary gauge choices). We consider the case of $q/p$ flux quantum per
square, where $q$ and $p$ are primitive integers. 
In this case, the original tight binding band (for zero field) is split into $p$ Landau subbands in the absence of the random potential.
The second term represents a random onsite potential. We take $\epsilon_i$
to be uniformly distributed between $-w$ and $w$, measured in units of the magnitude of
the nearest neighbor hopping matrix element, which is set to be 1 throughout
this paper.

The special case of the model (\ref{ham}) with near neighbor hopping only 
and $q=1$ has been studied in some detail 
before.\cite{yang,liu,xie,sheng,sheng2,ando,tan}
It is found in this case that
for weak randomness, the extended
states in most Landau subbands form a critical energy with total
Chern number\cite{chernnote} (TCN)
$+1$, while those in the central subband (odd $p$) or two central subbands
(even $p$) split into two critical energies with negative TCNs.
The total Chern number of the entire band is zero, as the filled band has
zero Hall conductance.
Increasing randomness strength merges critical energies with TCN $+1$ and
those with negative TCNs, with
those closer to the band center disappearing at lower randomness.\cite{ando,liu,sheng}
Simultaneously, it is found that the critical energies near the band edge float away from the
centers of their corresponding Landau subbands towards the center
of the whole band,\cite{yang} consistent with the ``floating up"
picture\cite{kh,laughlin}
in continuum systems where there is no critical energy with negative
TCN. Such behavior is understood heuristically as due to
an '`attraction" between states or critical energies with 
Chern numbers of opposite signs.\cite{yang}

This special case discussed above
has the very unique feature that in the absence of
randomness all subbands carry Chern number $+1$ except for the central
subband(s). In experiments however, the field strength is varied continuously.
Also restricting the model to near neighbor hopping only endows the model with 
particle-hole symmetry, which is also absent in real systems\cite{noteadd}.
For general field strength commensurate with the lattice,
i.e., $q/p$ flux quantum per plaquette, 
the Chern numbers of subbands follow some nontrivial
pattern and subbands carrying Chern numbers of opposite signs intertwine with
each other. It is unclear how the extended states behave upon introducing
randomness in this case.
We address this issue numerically in the present paper,
and without losing generality, we assume $q/p<1/2$.

To study the
localization property of this model, we use the Chern number 
approach,\cite{tknn,niu,arovas,huo} which is briefly reviewed below.
The Hall conductance of an individual eigenstate $|m\rangle$ can be
obtained easily using the Kubo formula:
\begin{eqnarray*}
\nonumber
\sigma_{xy}^{m}={ie^2\hbar\over A}\sum_{n\ne m}{\langle m|v_y|n\rangle
\langle n|v_x|m\rangle-\langle m|v_x|n\rangle\langle n|v_y|m\rangle\over
(E_n-E_m)^2},\nonumber
\end{eqnarray*}
where $A$ is the area of the system, $v_x$ and $v_y$ are the velocity operators
in the $x$ and $y$ directions respectively. For a finite system
with the geometry of a parallelogram with periodic boundary conditions
(torus geometry), $\sigma_{xy}^m$ depends on the two boundary
condition phases $\phi_1$ and $\phi_2$. As shown by Niu {\em et al.},
the boundary condition averaged Hall conductance takes the
form\cite{niu}
\begin{equation}
\langle\sigma_{xy}^m\rangle={1\over 4\pi^2}\int{d\phi_1d\phi_2\sigma_{xy}^{m}
(\phi_1,\phi_2)}=C(m)e^2/h,
\end{equation}
where $C(m)$ is an integer called the Chern number of the state $|m\rangle$.
States with nonzero Chern numbers carry Hall
current and are necessarily
extended states\cite{arovas,huo}. Thus by numerically diagonalizing
the Hamiltonian on a grid of $\phi_1$ and $\phi_2$, and calculating the
Chern numbers by converting the integral in
(2) to a sum over grid points, we are able to identify
extended states unambiguously. Depending on system size, the number of grids
used in numerical calculations range from $25 \times 25$ to $50 \times 50$.

\section{Numerical results}
\label{numer}

In this section we
present our numerical results. We start by discussing the
case when there is only nearest neighbor hopping, and no disorder.
In this case we have an exact particle-hole symmetry, and the quantized
Hall conductance of each of the $p$ Landau subbands (which is also the Chern
number of this subband) is known exactly. For the $t$th subband, the quantized
Hall conductance (in unit of $e^2/h$) $\sigma_t$ when it is occupied 
{\em together
with all the subbands below it}, satisfies the Diophantine
equation\cite{macd,tknn}
\begin{equation}
t=\sigma_tq+sp,
\label{di}
\end{equation}
where $s$ is an integer that gives rise to the 
integer $\sigma_t$ with the smallest
magnitude satisfying Eq. (\ref{di}).\cite{tknn,mouche} 
The quantized Hall conductance of the $t$th subband itself, 
or its Chern number, is therefore
\begin{equation}
C_t=\sigma_t-\sigma_{t-1}.
\end{equation}

In the following we show that the solutions of Eq. (\ref{di}) (together with
the condition below it) has a very non-trivial pattern in it. We start by 
discussing the case $t=mq$, where $m$ is an integer. In this case 
Eq. (\ref{di}) reduces to
\begin{equation}
\sigma_{mq}q=mq-sp.
\end{equation}
For $m<p/2$, the solution with $\sigma_{mq}$ having smallest magnitude is
realized by choosing $s=0$, and
\begin{equation}
\sigma_{mq}=m.
\end{equation}
Therefore for $m<p/2$,
\begin{equation}
\sum_{i=(m-1)q+1}^{mq}{C_i}=\sigma_{mq}-\sigma_{(m-1)q}=1,
\end{equation}
implying that in the lower half of the entire band, {\em subbands form 
groups of $q$ subbands with total Chern number 1}. From particle-hole symmetry
we know that the same happens in the upper half of the entire band as well;
and the remaining $M={\rm mod}(p, 2q)$ subbands at the center of the entire
band
therefore form a special group carrying negative total Chern numbers, so that
the total Chern number for the entire band is zero.

The Chern numbers of individual subbands, however, depends sensitively 
on $q$ and $p$, and oscillates between positive and negative integers.
Therefore in the absence of disorder, the quantized Hall conductance does not
have a simple monotonous dependence on the filling factor, even if the Fermi
energy is far away from the band center. Also in this case it is not clear
how current carrying states move and merge upon introducing disorder, based
on the heuristic argument that ``states carrying Chern numbers with opposite
signs attract each other".\cite{yang}

In the following we show that as randomness is introduced, it first merges 
the current carrying states {\em in the same group}. More specifically, 
current carrying states in groups away from the band center merge together 
and form  critical energies carrying total Chern number $+1$, while 
those in the group at the band center form two critical energies carrying
equal amount of negative Chern numbers. This is because the energy gaps 
separating subbands within the same group are much smaller than those
separating neighboring groups, therefore the density of states as well as
current carrying states in the same
group merge together first. At this point the configuration of current 
carrying critical energies become the same as that of the cases with $q=1$:
There are critical energies with TCN $+1$ away from the band center, and 
two critical energies carrying negative TCN near the band center. Further 
increasing the randomness strength will merge the remaining critical energies,
with the ones closer to the band center disappearing 
earlier.\cite{liu,xie,sheng,sheng2}

In the next two subsections we present numerical results for cases we have studied, to
demonstrate the above observations.   
It should be borne in mind that the numerical results presented are for finite sized lattices,
which may differ somewhat from the results of lattices in the thermodynamic limit (e.g., some critical
energies which may be disctinct in the latter case may not be resolved for our 
sizes). However,
we wish to emphasize that we have taken care to make sure that all the relevant splittings are
observable for our sizes, except possibly a small splitting of the central band in the particle-hole
symmetric case.

\subsection{Cases with nearest neighbor hopping only}

We first present our results for cases where there is 
nearest neighbor hopping only. In these cases we have particle-hole symmetry
even in the presence of random potential, {\em after} physical quantities 
are averaged over all possible disorder configurations. The number of 
different disorder configurations (or ``samples") we 
explore numerically range from 200 to 800.
The special case of $q/p=2/5$ has
already been studied by Tan,\cite{tan} the results obtained by him agree with our
general observations above.

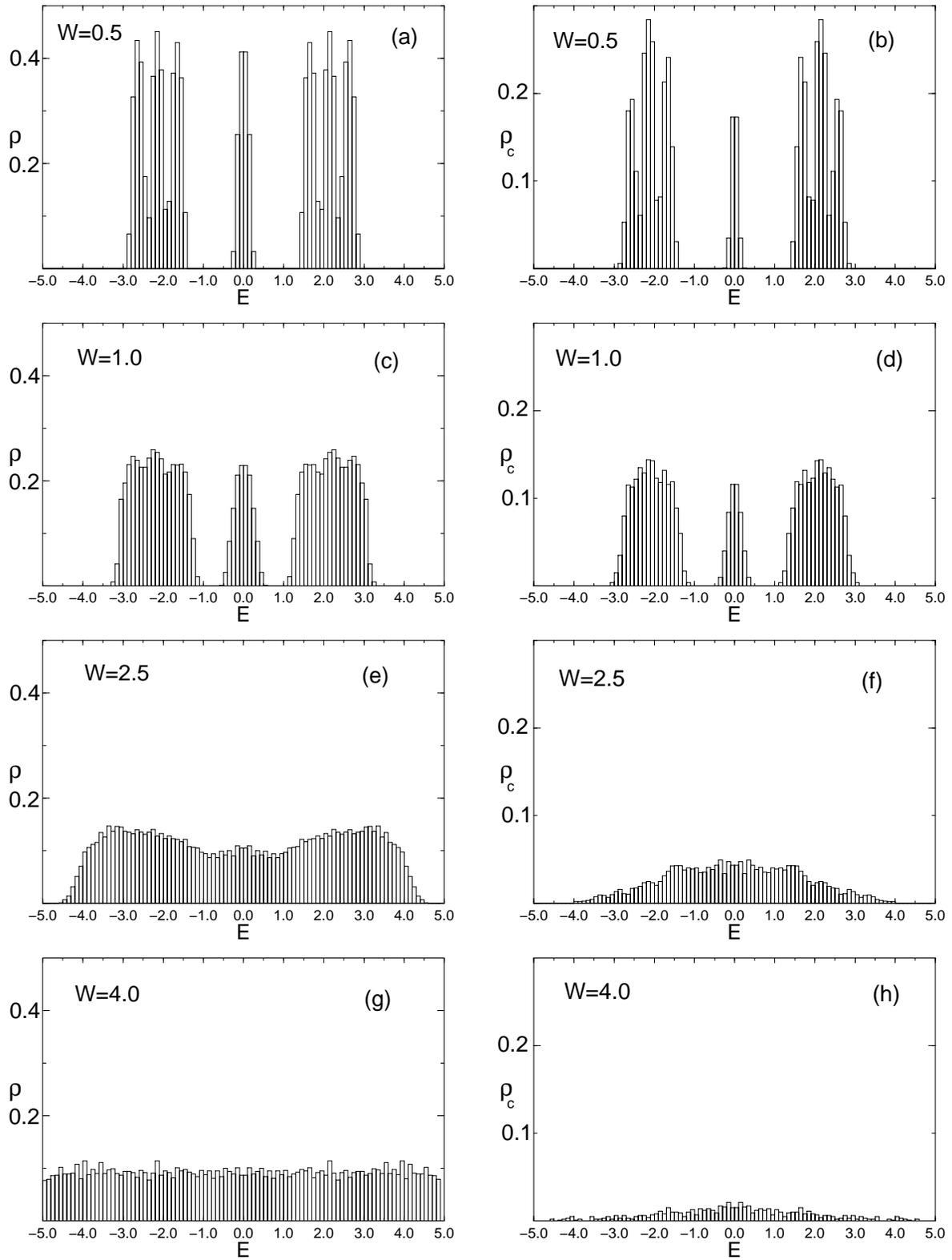
\begin{figure}
\caption{Total density of states per unit area ($\rho$, left panels), and
density of current carrying states ($\rho_c$, right panels), for systems with
$3/7$ flux quantum per plaquette, size $7\times 7$, and different disorder
strengths ($w$).
}
\label{37}
\end{figure}

{\em $q/p=3/7$}. In this case we have 7 subbands in the absence of disorder,
and the Chern numbers of the subbands are $(-2, 5, -2), (-2), (-2, 5, -2)$,
obtained from solving Eq. (\ref{di}) with the condition below it,
with ones in the same parentheses belong to the same group.
In Fig.\ref{37} we show how the
disorder averaged total density of states ($\rho$), as well
as density of current carrying states ($\rho_c$), evolve as the randomness
strength $w$ increases, for systems of square geometry and size $7\times 7$.
We see for weak randomness of $w=0.5$, the tails of $\rho$ of different 
subbands in the same group have just started to overlap, and there are still 
well defined peaks of $\rho_c$ for each 
individual subband. In the thermodynamic limit, current carrying extended
states are expected to exist 
only at individual critical energies;\cite{bodoreview}
in finite size systems however, $\rho_c$ has finite width around these 
critical energies as states with localization length bigger than the 
system size appear extended. As $w$ is increased to $1.0$, subbands in the same 
group have merged together, so have $\rho_c$, while there are still gaps 
separating different groups. Further increasing $w$ to $2.5$, 
$\rho$ of the 3 different group start to merge; the current carrying states
in the central group (which consists of only a single subband in this case)
split into two peaks (that become sharp critical energies each carrying
TCN -1 in the thermodynamic limit), 
moving toward the band edges to meet the current carrying states 
in the side groups, which form two critical energies each carrying TCN +1 in
the thermodynamic limit.
At this point, as far as the behavior of current carrying states is 
concerned, the situation is identical to that of the case with $q/p=1/3$ 
which has been studied closely earlier,\cite{yang,yang1}
even though initially we have 7 subbands in this case.
Further increasing $w$ to $4.0$, the critical energies carrying opposite 
TCN have merged and disappeared, and there are no current carrying states 
anywhere in the thermodynamic limit; the system is in an insulating state no
matter where the Fermi energy is. In a finite size system we see very 
low $\rho_c$ which is expected to go to zero rapidly as system size
increases.

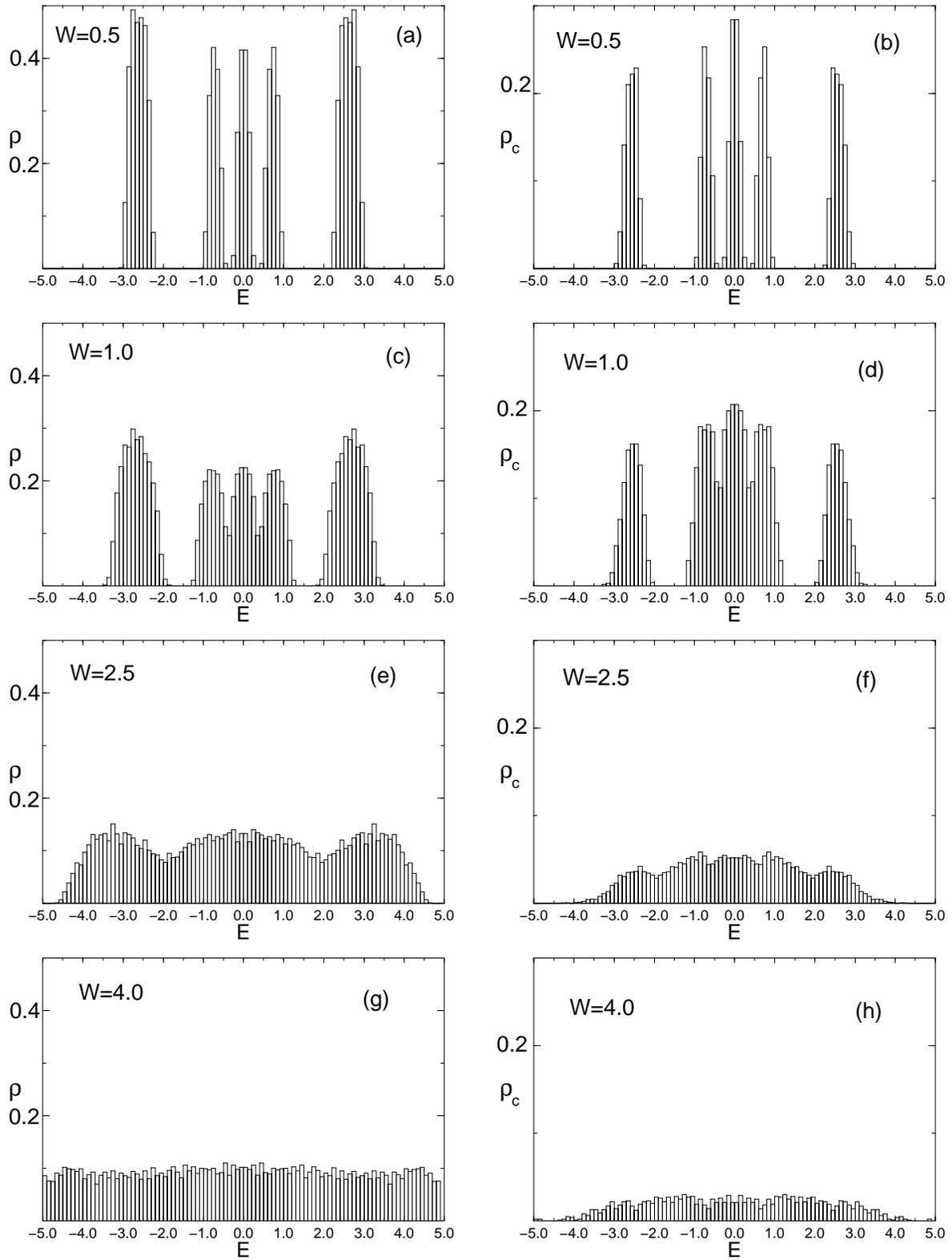
\begin{figure}
\caption{
Same as in Figure 1, for systems with
$2/7$ flux quantum per plaquette, size $7\times 7$.
}
\label{27}
\end{figure}

{\em $q/p=2/7$}. In this case we again
have 7 subbands in the absence of disorder,
and the Chern numbers of the subbands are $(-3, 4), (-3, 4, -3), (4, -3)$,
again
with the ones in the same parentheses belong to the same group.
We again study systems with size $7\times 7$, and show in Fig.\ref{27} how
$\rho$ and $\rho_c$ evolve with $w$ in this case.
For $w=0.5$, the subbands in the two side groups have already merged together,
while the tails of $\rho$ of the central group (with 3 subbands) barely start
to overlap. Increasing $w$ to $1.0$, the 3 subbands in the central group now
have significant overlap, while the gaps separating different groups remain.
As in the previous case, further increasing $w$ to $2.5$,
$\rho$ of the 3 different group start to merge, and 
he current carrying states
in the central group split into two peaks, 
moving toward the band edges to meet the current carrying states
in the side groups. Further increasing $w$ to $4.0$, all critical energies 
have merged and disappeared, and we see a very low $\rho_c$.

$q/p=3/8$. In this case we have 8 subbands in the absence of disorder,
and the Chern numbers 
of the subbands are $(3, -5, 3), (-1, -1), (3, -5, 3)$.\cite{note38}
We have studied systems with size $8\times 8$.
The situation is quite similar to the previous two cases.
For $w=0.3$, $\rho$ of the subbands in the same group have just started to 
overlap. As $w$ is increased to $1.0$, subbands in the same
group have merged together, so have $\rho_c$, while there are still gaps
separating different groups. Further increasing $w$ to $2.2$,
$\rho$ of the 3 different group start to merge; the current carrying states
in the central group 
split into two peaks, 
moving toward the band edges to meet the current carrying states
in the side groups, which form two critical energies each carrying TCN +1 in
the thermodynamic limit.
Further increasing $w$ to $3.0$, the critical energies carrying opposite TCN
have just merged and disappeared.

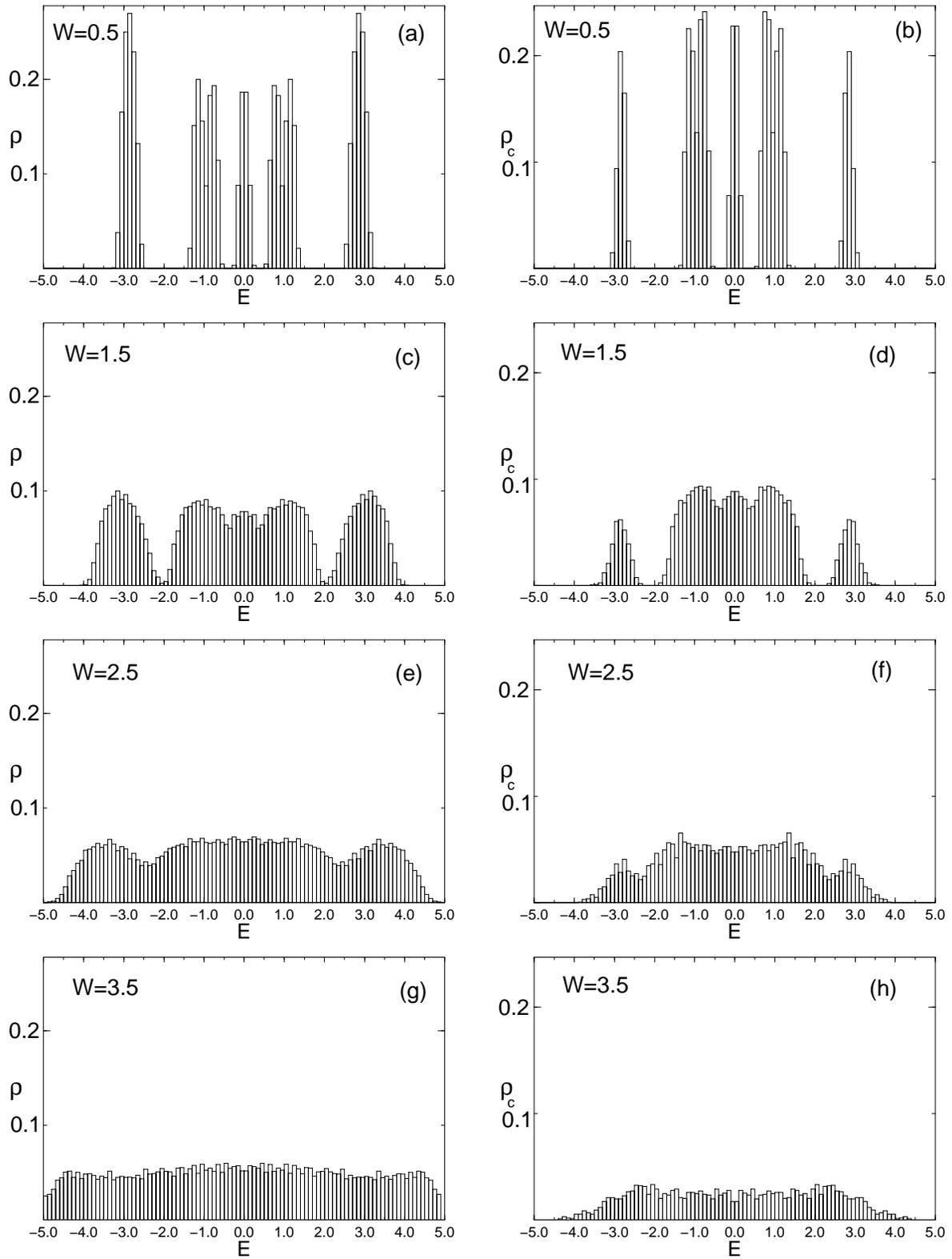
\begin{figure}
\caption{
Same as in Figure 1,
for systems with
$2/9$ flux quantum per plaquette, size $9\times 9$.
}
\label{29}
\end{figure}

$q/p=2/9$. In this case we have 9 subbands in the absence of disorder,
and the Chern numbers
of the subbands are $(-4, 5), (-4, 5), (-4), (5, -4), (5, -4)$.
The dependence of $\rho$ and $\rho_c$ on $w$ for systems of size
$9\times9$ are shown in Fig. \ref{29}.
For $w=0.5$, $\rho$ and $\rho_c$ of groups 1 and 5 have fully merged 
already, and those in groups 2 and 4 have started to merge, although 
we still have separate peaks there. Gaps between groups are quite 
pronounced at this point. Increasing $w$ to 1.5, the subbands in groups 2 and
4 have also fully merged; and they have also started to merge with the 
central group (consisting of only 1 subband). At this point the situation has
become identical to that of $q/p=1/5$: we have 4 critical energies each 
carrying TCN +1 away from the band center, while the current carrying states
at band center has TCN -4. Further increasing $w$ to 2.5, we find the 
current carrying states from the central group have split into two critical
energies, and they in turn have already merged with the current carrying states
from groups 2 and 4; there remain 4 critical energies as indicated by the 4
peaks in $\rho_c$; 2 of them carry TCN +1, from the edge groups; the other 2
closer to the band center carry TCN -1. All the remaining critical energies
have merged and disappeared as $w$ is increased to 3.5.

\subsection{Cases with next nearest neighbor hopping}

In this subsection we present data for cases where there is next nearest 
hopping, with the magnitude of the hopping matrix element $t'$, in unit of the
nearest neighbor hopping. In the presence of $t'$, the particle-hole
symmetry present in the model with nearest neighbor hopping only is broken. Although Eq. (\ref{di}) is still satisfied in this case,
generically the magnitude of
$\sigma_t$ is not necessarily 
minimized (by choosing the appropriate $s$). However if $t'$ is small enough, 
and if the subband gaps are not closed as $t'$ is increased from 0, the 
Chern number of each subband remains the same as that of $t'=0$.

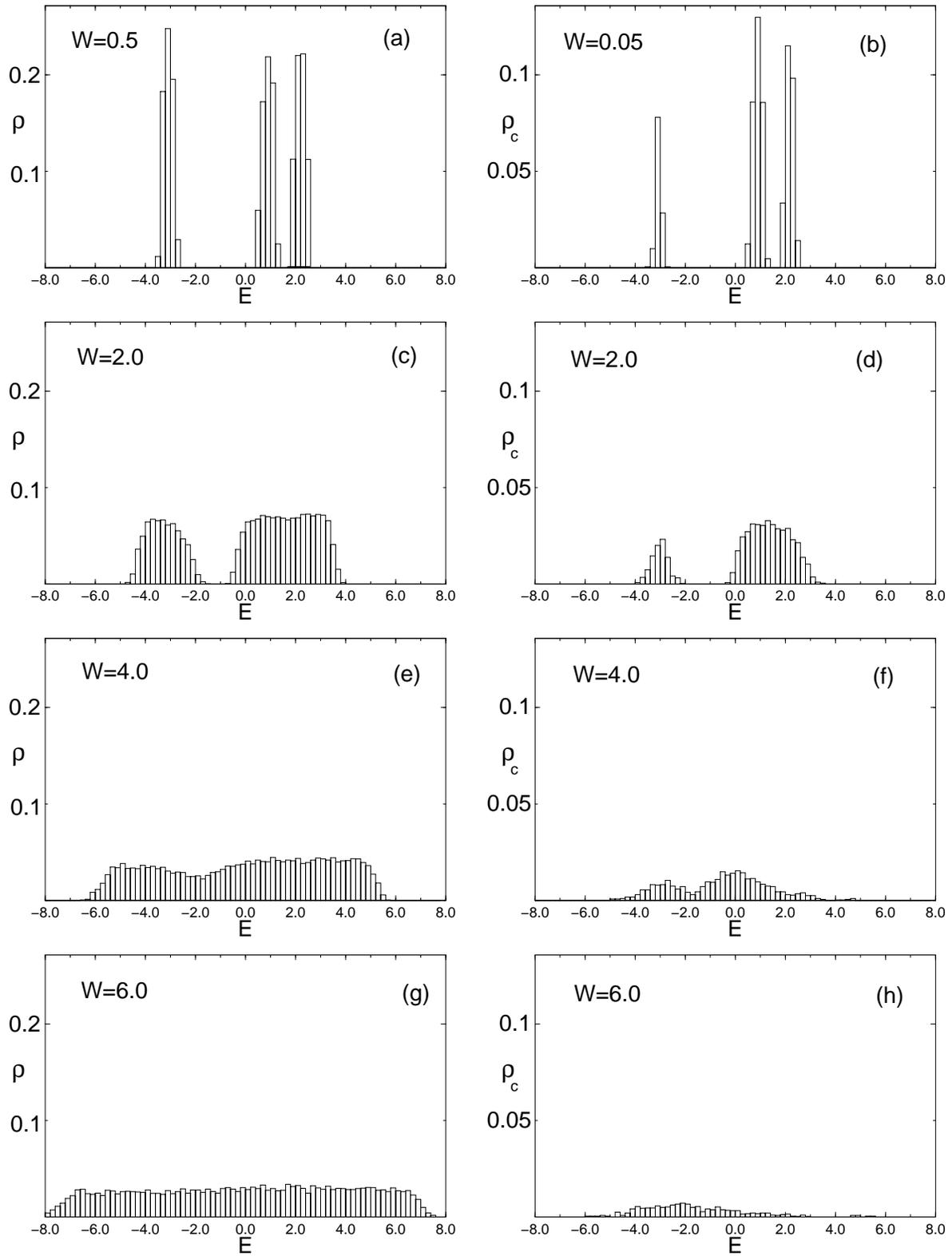
\begin{figure}
\caption{
Same as in Figure 1,
for systems with
$1/3$ flux quantum per plaquette, magnitude of next nearest neighbor hopping 
$t'=0.5$, size $9\times 9$.
}
\label{13}
\end{figure}

$q/p=1/3, t'=0.5$. In this case we have 3 subbands. The Hall conductance 
(Chern number) of each subband is +1, +1, and -2. This is {\em different}
from the case with $t'=0$, where the Chern numbers are +1, -2, +1. We find
the next nearest neighbor hopping term breaks the particle-hole symmetry,
and moves the negative Chern numbers to higher subbands. In Fig. \ref{13}
we show how $\rho$ and $\rho_c$ for systems of size $9\times 9$
evolve with $w$.
For weak randomness $w=0.5$, the gaps separating different subbands are
still finite. Increasing $w$ to 2.0, the central and upper subbands have
merged, and the current carrying states in both subbands have also merged
to form one critical energy represented by a single peak, carrying total
Chern number -1. Further increasing $w$ to 4.0, the lower subband also starts
to merge with the other subbands, and the two remaining critical energies
carrying Chern numbers +1 and -1 respectively start to move together.
When $w$ is increased to 6.0, all the subbands have merged together 
completely, and the two critical energies have also merged and disappeared. 

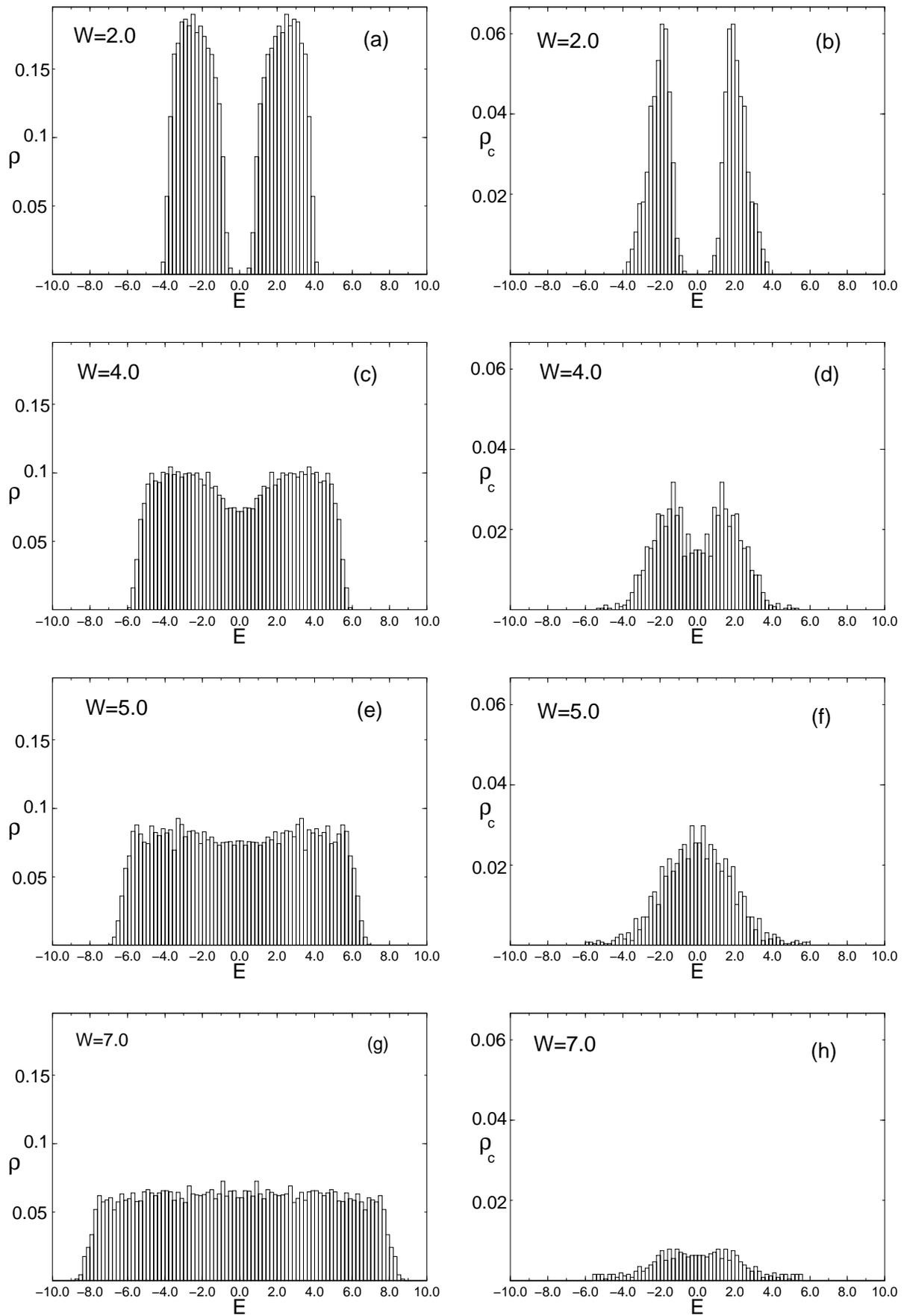
\begin{figure}
\caption{
Same as in Figure 1,
for systems with
$1/2$ flux quantum per plaquette, magnitude of next nearest neighbor hopping 
$t'=0.5$, size $8\times 8$.
}
\label{12}
\end{figure}

$q/p=1/2, t'=0.5$. This is a somewhat special case, because with $t'=0$, the
system actually respects time-reversal symmetry (since the nearest neighbor
hopping matrix elements can be chosen to be real), and the Hall conductances
(and Chern numbers) are zero everywhere. Related to this is the fact that the
two subbands are degenerate at two points in the Brillouin zone in this case.
In the presence of a nonzero $t'$, a gap separating the two subbands is opened up, and the
Chern numbers of the lower and upper subbands become 1 and -1 respectively.
In Fig. \ref{12} we show results of numerical studies when random potential is
added to the system, for systems with size $8\times 8$. 
Previously Ludwig {\em et al.}\cite{ludwig} 
studied this case 
analytically. We see when $w=2.0$, there is still a gap separating the two
subbands, and the current carrying states form two
critical energies represented by the peaks in $\rho_c$. Increasing $w$ to
4.0, the two subbands have started to merge, and the two critical energies 
start to move closer together toward the band center. Further 
increasing $w$ to 5.0, the two subbands have merged completely, and the 
two critical energies 
have almost merged.  Further 
increasing $w$ to 7.0, the two critical energies 
have merged and disappeared, and there are very few current carrying states
left in such a finite size system.
 
\section{summary and discussion}
\label{sum}

In this paper we have studied numerically how current carrying states 
disappear, and the system undergoes phase transitions from a quantum Hall
state to an insulating state as disorder strength increases, in tight binding
lattice models with non-interacting electrons. In cases with
nearest neighbor hopping only (with particle-hole symmetry), 
we find the following 
generic pattern: 
Subbands away from the center of the band form groups with
$q$ subbands, and the total Chern number
of such groups is $+1$; 
while subbands near the band center for a special group
with $M={\rm mod}(p, 2q)$ subbands.
In the presence of relatively weak randomness, groups of
critical
energies away from the band center merge together and form critical energies
carrying Hall conductance (or total Chern number) 1; the current carrying
states near the band 
center form (within our resolution) two critical energies carrying negative 
Chern numbers. Further increasing randomness strength, the critical energies 
carrying negative
Chern numbers move toward the band edge and merge with the critical energies
carrying total Chern number 1, and eventually all critical energies and 
current carrying states disappear in the thermodynamic limit, and the system
becomes insulating no matter where the Fermi energy is.
Adding a next nearest neighbor hopping term breaks the particle-hole symmetry.
We find this tends to push subbands carrying negative Chern numbers to higher
energies. However as long as the Fermi energy is near the bottom of the band,
the way critical energies carrying positive Chern numbers are killed is 
similar: they are killed by critical energies carrying negative Chern numbers
coming down from high energy.\cite{holenote}

Our results suggest that the non-trivial structure in the dependence of Hall 
conductance on electron density in the presence of a periodic potential is
quite vulnerable against randomness; in order to see such structures 
experimentally,\cite{klitzing} one must control the amount of randomness
carefully so that subbands in the same group (which are separated by very small
gaps) are not merged together by randomness.
As the randomness becomes strong enough to merge subbands in the same group,
there remain only critical energies carrying Hall conductance 1 except near the
band center. Thus the situation is essentially the same as 
without periodic potential, i.e., there is no oscillation 
in the dependence of the Hall conductance on electron density.

In the one-band tight binding model studied here and elsewhere, all current
carrying extended states disappear at finite randomness strength for a fixed
strength of magnetic field. Based on this observation, 
Xie {\em et al.}\cite{xie} proposed that in the {\em continuum} system,
instead of floating to infinite energy in the limit $B\rightarrow 0$, 
all extended states disappear at {\em finite} $B$ for a given disorder 
strength. In the following we show that extrapolating the lattice model to the
continuum limit is quite subtle, and one needs to be extremely careful in 
interpreting the implications of results of lattice model 
on the continuum system. For simplicity we consider the case $q=1$; 
extension to cases with $q>1$ 
in the following discussion is straightforward.\cite{sondhi}

Without randomness, there are two energy scales in the lattice model for a 
given strength of magnetic field, namely, the bandwidth which is of order
$t$ (the hopping matrix element), and the Landau level spacing 
$\hbar\omega_c=\hbar eB/m^*c\propto t/p$, 
where $m^*$ is the effective mass at the bottom of the
band. There are also two length scales, the lattice spacing $a$ and the
magnetic length $\ell=\sqrt{\hbar c/eB}\propto a\sqrt{p}$.
In the continuum system however, there is only one energy scale, namely the
Landau level spacing $\hbar\omega_c$, and one length scale, the magnetic
length $\ell$; the band width is infinite, and the lattice spacing is zero.
Thus to reach the continuum limit in the tight binding lattice model, 
one must send both $t$ and $p$ to infinity, with the ratio $t/p$ remain a 
constant; in the mean time $a$ should be sent to zero in such a way that
$a\sqrt{p}$ remains a constant.

The random potential
introduces another energy scale.\cite{notelength}
Naively, the new scale is $w$; what
has been shown in lattice models is that the critical $w_c$ above which no
extended states exist is finite {\em in units of t},\cite{liu,xie,sheng}
and $w_c$ goes to zero as $p$ goes to infinity, again {\em in units of t}.
In the continuum model however, the real new energy scale
introduced by the random potential is the amount of broadening 
of a Landau level, $\delta E$. The important question in the continuum 
model is {\em in units of
$\hbar\omega_c$},  
how much $\delta E$ does one need to kill all the current
carrying extended states, or what is the ratio $\delta E_c/\hbar\omega_c$? 
If the answer is finite, then for a given 
disorder strength, one needs a finite critical magnetic field in the
continuum for extended states to exist, as suggested 
in Refs. \onlinecite{liu,xie}; while if the answer is infinite, then
the extended state must exist no matter how small $B$ is, and float to 
infinite energy only in the limit $B\rightarrow 0$, as originally 
proposed by Khmelnitskii\cite{kh} and Laughlin.\cite{laughlin}
In the lattice model with large $p$, we have $\delta E\propto w/\sqrt{p}$.
This is because the linear size of
the wave function in a given Landau level is 
$\ell$, which means the number of lattice sites it occupies 
is proportional to $\ell^2\propto p$; since the random potential on 
different sites are uncorrelated, the shift on the energy is typically of order
$w/\sqrt{p}$, the magnitude of the {\em averaged} potential in that
area. Therefore we find
\begin{equation}
{\delta E_c\over \hbar\omega_c}\propto{w_c/\sqrt{p}\over t/p}
={w_c\sqrt{p}\over t}.
\label{ratio}
\end{equation}
Let us assume 
\begin{equation}
w_c\propto1/p^{\alpha}
\label{asp}
\end{equation}
in the limit $ p\rightarrow\infty$.
Since $\delta E$ must be at least
comparable to $\hbar\omega_c$ for current carrying 
states to disappear, from Eq. (\ref{ratio}) we must have $\alpha \le 1/2$.
If $\alpha=1/2$, the ratio in Eq. (\ref{ratio}) is finite as 
$ p\rightarrow\infty$; while if $\alpha < 1/2$, or $\alpha=1/2$ but 
there is logarithmic corrections to Eq. (\ref{asp}), the ratio in Eq. 
(\ref{ratio}) 
becomes infinite as 
$ p\rightarrow\infty$; the situation becomes qualitatively different.
It was shown in Ref. \onlinecite{sheng} that the data for $16 \le p \le 384$
may be fit reasonably well for $\alpha=1/2$.
Since the crucial question is whether $\alpha$ is {\em exactly} $1/2$ or not,
we feel it is probably difficult to resolve this issue numerically, and
analytic input is needed here.

\acknowledgements
The authors are particularly grateful to Allan MacDonald and Shivaji Sondhi
for important conversations. They have also benefited from
useful discussions and correspondences
with Mike Hilke,
Danny Shahar,
Dong-Ning Sheng, 
S.-H. Song,
David Thouless, Dan Tsui, Zheng-Yu Weng, and Xin-Cheng Xie. 
This work was supported by NSF grant DMR-9400362 (at Princeton), and the
Sherman Fairchild Foundation (at Caltech).
R. N. B. acknowledges the hospitality of the Aspen Center for Physics
while this manuscript was being written.

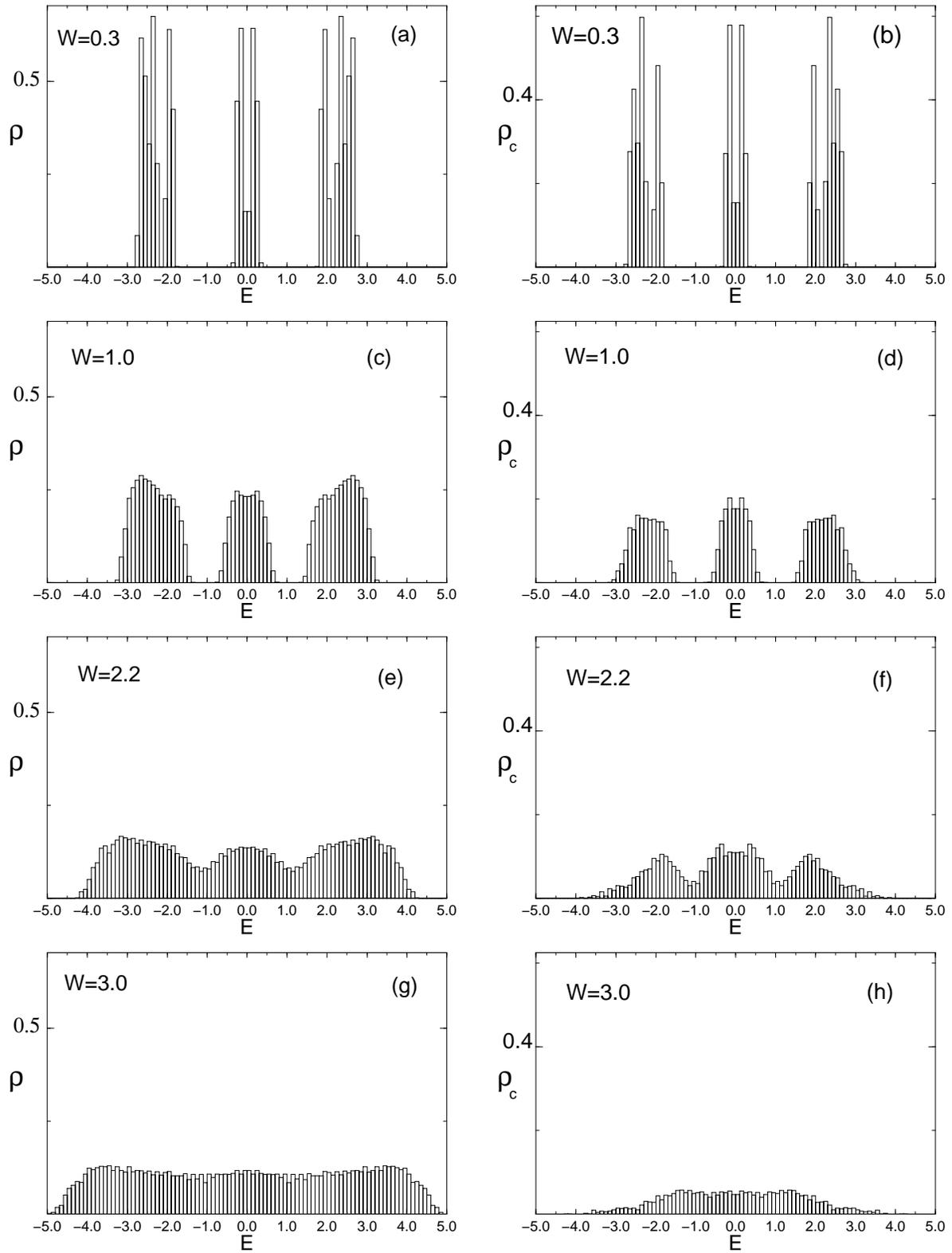
\begin{figure}
\caption{
Same as in Figure 1,
for systems with
$3/8$ flux quantum per plaquette, size $8\times 8$.
}
\label{38}
\end{figure}

\begin{references}

\bibitem{pg} For reviews, see {\em The Quantum Hall Effect}, edited by
R. E. Prange and S. M. Girvin (Springer-Verlag, New York, 1990);
{\em Perspectives in Quantum Hall Effect}, edited by S. Das Sarma and A.
Pinczuk (Wiley, New York, 1997).

\bibitem{bodoreview} For a review, see
B. Huckestein, Rev. Mod. Phys. {\bf 67}, 357 (1995).

\bibitem{noteadd} The critical energies are exactly
at the centers of Landau bands
only for electron-hole (e-h) symmetric models. In real semiconductor
systems, this is 
only approximately 
true, with the e-h 
symmetry broken both by (i) a non-symmetric random potential, and (ii) mixing
between different Landau levels.

\bibitem{gangof4} E. Abrahams, P. W. Anderson, D. C. Licciardello and
T. V. Ramakrishnan,
Phys. Rev. Lett. {\bf 42}, 673 (1979).

\bibitem{note} There is experimental evidence (S. V. Kravchenko {\em et al.},
Phys. Rev. B {\bf 50}, 8039 (1994)) that a metallic phase may be 
stabilized by strong electron-electron interactions at $B=0$. We do not
consider this possibility here.

\bibitem{jiang} H. W. Jiang {\em et al.}, Phys. Rev. Lett. {\bf 71},
1439 (1993).

\bibitem{wang} T. Wang {\em et al.}, Phys. Rev. Lett. {\bf 72}, 709 (1994).

\bibitem{shahar} D. Shahar, D. C. Tsui and J. E. Cunningham, Phys. Rev. B {\bf
52}, R14372 (1995).

\bibitem{song} S.-H. Song, D. Shahar, D. C. Tsui, Y. H. Xie and
D. Monroe, Phys. Rev. Lett. {\bf 78}, (1997).

\bibitem{kh} D. E. Khmelnitskii, Phys. Lett. A {\bf 106}, 182 (1984).

\bibitem{laughlin} 
R. B. Laughlin, Phys. Rev. Lett. {\bf 52}, 2304 (1984).

\bibitem{klz} S. Kivelson, D. H. Lee and S. C. Zhang, Phys. Rev. B {\bf 46},
2223 (1992).

\bibitem{shashkin} A. A. Shashkin. V. T. Dolgopolov and G. V. Kravchenko.
Phys. Rev. B {\bf 49}, 14486 (1994).

\bibitem{glozman} I. Glozman, C. E. Johnson and H. W. Jiang, Phys. Rev. Lett.
{\bf 74}, 594 (1995).

\bibitem{krav} S.V. Kravchenko {\em et al.}, Phys. Rev. Lett. {\bf 75},
910
(1995).

\bibitem{furneaux} J.E. Furneaux {\em et al.}, Phys. Rev. B {\bf 51},
17227 (1995).

\bibitem{a} T. Ando, J. Phys. Soc. Jpn. {\bf 53}, 3126 (1984).

\bibitem{ld}
D. Liu and S. Das Sarma, Phys. Rev. B {\bf 49},  2677 (1994).

\bibitem{shahbazyan} T. V. Shahbazyan and M. E. Raikh, Phys. Rev. Lett.
{\bf 75}, 304 (1995).

\bibitem{yang} K. Yang and R. N. Bhatt, Phys. Rev. Lett. {\bf 76}, 1316 (1996).

\bibitem{gramada} A. Gramada and M. E. Raikh, Phys. Rev. B {\bf 54}, 1928
(1996).

\bibitem{haldane} F. D. M. Haldane and K. Yang, Phys. Rev. Lett. {\bf 78},
298 (1997).

\bibitem{liu} D. Z. Liu, X. C. Xie and Q. Niu,
Phys. Rev. Lett. {\bf 76}, 975 (1996).

\bibitem{xie}X. C. Xie, D. Z. Liu, B. Sundaram and Q. Niu,
Phys. Rev. B {\bf 54}, 4966 (1996).

\bibitem{sheng} D. N. Sheng and
Z. Y. Weng, Phys. Rev. Lett. {\bf 78}, 318 (1997).

\bibitem{sheng2} D. N. Sheng and
Z. Y. Weng, Phys. Rev. Lett. {\bf 80}, 580 (1998).

\bibitem{ando} T. Ando, Phys. Rev. B {\bf 40}, 5325 (1989).

\bibitem{tan} Y. Tan, J. Phys. Condens. Matter {\bf 6}, 7941 (1994).


\bibitem{chernnote} To be defined below. See also Refs. 
[\onlinecite{tknn,niu}].

\bibitem{tknn} D. J. Thouless, M. Kohmoto, M. P. Nightingale, and M. den
Nijs,
Phys. Rev. Lett. {\bf 49}, 405 (1982).

\bibitem{niu} Q. Niu, D. J. Thouless and Y. S. Wu, Phys. Rev. B {\bf 31},
3372 (1985).

\bibitem{arovas} D. P. Arovas {\em et al.},
Phys. Rev. Lett. {\bf 60}, 619 (1988).

\bibitem{huo} Y. Huo and R.N. Bhatt, Phys. Rev. Lett. {\bf 68}, 1375 (1992).

\bibitem{macd} A. H. MacDonald, Phys. Rev. B {\bf 28}, 6713 (1983).
See also A. H. MacDonald, in {\em Quantum Coherence in Mesoscopic
Systems}, edited by B. Kramer, Plenum Press, New York, 1991.

\bibitem{mouche} P. von Mouche, Comm. Math. Phys. {\bf 122}, 23 (1989).

\bibitem{note38} Strictly speaking in this case (and all other cases with 
even $p$), the Chern numbers of the two central bands are not well defined, as
there is no gap separating them; only the sum of their Chern numbers are
well defined. However a gap may be opened by introducing a very small next
neighbor hopping term to make the Chern numbers well defined.

\bibitem{yang1} K. Yang, D. Shahar, R. N. Bhatt, D. C. Tsui and M. Shayegan,
preprint cond-mat/9805341.

\bibitem{ludwig} A. W. W. Ludwig, M. P. A. Fisher, R. Shankar and
G. Grinstein, Phys. Rev. B {\bf 50}, 7526 (1994).

\bibitem{holenote} The situation however may be different when the carriers
are holes instead of electrons; further investigation is underway. 

\bibitem{klitzing} For early experimental attempts in studying the IQHE in the
presence of a periodic potential, see R. R. Gerhardts,
D. Weiss, and K. v. Klitzing, Phys. Rev. Lett. {\bf 62}, 1173 (1989);
R. W. Winkler, J. P. Kotthaus, and K. Ploog,
Phys. Rev. Lett. {\bf 62}, 1177 (1989).

\bibitem{sondhi} We thank S. L. Sondhi for extensive discussions on the 
following points.

\bibitem{notelength} For the uncorrelated random onsite potential studied 
here (which maps onto
Gaussian white noise potential with zero correlation length in the
continuum limit), no new length scale is introduced, although it would be
interesting to study random potential with finite correlation length as well.

\end{references}
\end{document}